\documentclass[aps,prb,reprint,floatfix]{revtex4-2}

\usepackage{graphicx}
\usepackage{xcolor}
\usepackage[colorlinks=true, urlcolor=blue, citecolor=blue, linkcolor=blue]{hyperref}
\usepackage{dcolumn}

\begin{document}

\title[Resistivity and Hall effect of HgBa$_2$CaCu$_2$O$_6$ thin films]{Resistivity, Hall effect, and anisotropic superconducting coherence lengths of HgBa$_2$CaCu$_2$O$_{6+\delta}$ thin films with different morphology}

\author{H.~Richter}

\author{W.~Lang}
\email[Corresponding author: ]{wolfgang.lang@univie.ac.at}
\affiliation{University of Vienna, Faculty of Physics, Electronic Properties of Materials, Boltzmanngasse 5, A-1090, Wien, Austria}

\author{M.~Peruzzi}

\author{H.~Hattmansdorfer}
\affiliation{Johannes-Kepler-University Linz, Institute of Applied Physics, Altenbergerstrasse 69, A-4040 Linz, Austria}

\author{J.~H.~Durrell}
\altaffiliation{Permanent address: Engineering Department, University of Cambridge, Trumpington Street, Cambridge CB2 1PZ, United Kingdom}

\author{J.~D.~Pedarnig}
\affiliation{Johannes-Kepler-University Linz, Institute of Applied Physics, Altenbergerstrasse 69, A-4040 Linz, Austria}

\begin{abstract}
Thin films of the high-temperature superconductor
HgBa$_2$CaCu$_2$O$_{6+\delta}$ have been prepared on SrTiO$_3$ substrates by pulsed-laser deposition of precursor films and subsequent annealing in mercury-vapor atmosphere. The microstructural properties of such films can vary considerably and have been analyzed by x-ray diffraction and atomic force microscopy. Whereas the resistivity is significantly enhanced in samples with coarse-grained structure, the Hall effect shows little variation. This disparity is discussed based on models for transport properties in granular materials. We find that, despite of the morphological variation, all samples have similar superconducting properties. The critical temperatures $T_c \sim 121.2\,$K $\dots 122.0\,$K, resistivity, and Hall data indicate that the samples are optimally doped. The analyses of superconducting order parameter fluctuations in zero and finite magnetic fields yield the in-plane $\xi_{ab}(0) \sim 2.3\,$nm\,$\dots 2.8\,$nm and out-of-plane $\xi_{c}(0) \sim 0.17\,$nm\,$\dots 0.24\,$nm Ginzburg-Landau coherence lengths at zero temperature. Hall measurements provide estimates of carrier scattering defects in the normal state and vortex pinning properties in the superconducting state inside the grains.
\end{abstract}


\maketitle

\section{Introduction}

The mercury cuprates of the Hg-Ba-Ca-Cu-O family form a homologous series with the chemical formula HgBa$_2$Ca$_{n-1}$Cu$_n$O$_{2n+2+\delta}$ (HBCCO). The discovery
of high-temperature superconductivity in the $n=1$ compound
\cite{PUTI93} and the even higher transition temperatures $T_c=
120$~K in the $n=2$ \cite{PUTI93a} and $T_c= 135$~K in the $n=3$
material \cite{SCHI93}, respectively, has triggered enormous
research interest. The latter compound still holds the record for the highest critical temperature $T_c$ of any superconductor at ambient pressure and for a cuprate superconductor with $T_{c, onset}=164$\,K under quasi-hydrostatic pressure of 31\,GPa \cite{GAO94}. Mercury cuprates have been synthesized from $n = 1$ to 7 with $T_c$ raising with the number $n$ of neighboring CuO$_2$ layers up to a maximum at $n = 3$ and then decreasing
for $n > 3$ \cite{KUZE00}.

In contrast to their very promising superconducting properties, the mercury cuprates are hard to synthesize and handle due to the highly volatile and toxic nature of Hg and Hg-based compounds. For a possible effective use, but also for a measurement of the basic intrinsic properties, the fabrication of high-quality thin films is demanded. Several groups have succeeded in this task, mainly by using pulsed-laser deposition (PLD) of a precursor film with subsequent annealing in Hg and O$_2$ containing atmosphere \cite{WANG93,TSUE94,YUN95,YUN96}. Later it was demonstrated that HBCCO thin films can be grown on vicinal substrates in a well-oriented manner and form a roof-tile like structure that allows for measurements of in-plane and out-of-plane properties on the very same sample \cite{YUN00,OGAW04}.

As a consequence of the subtle preparation conditions, {the properties of HBCCO samples reported by various groups vary significantly. Investigations of the electrical transport properties in samples with different} structural properties have been rare. {It remains ambiguous whether such diversity stems from slightly different preparation conditions in individual laboratories or can occur also at supposedly identical fabrication procedures.} For instance, in polycrystalline samples of HgBa$_2$CaCu$_2$O$_{6+\delta}$ (Hg-1212) a room temperature variation of the electrical resistivity by a factor larger than two was observed between individual samples that were cut from the same ceramics but annealed for different time intervals \cite{HARR94b}. Surprisingly, the Hall effect of these samples was remarkably similar. On the other hand, both the resistivity and the Hall effect changed significantly in a Hg-1212 thin film after several annealing steps in different mercury and oxygen atmospheres \cite{SUN01}.

Several authors have investigated the electrical transport
properties, such as resistivity and the Hall effect, in HBCCO in detail, but without putting emphasis on possible variations between individual samples. Mostly, results of only a single sample were presented. In the mixed-state Hall effect, a double sign reversal, similar to that observed in other cuprate superconductors with high anisotropy, was reported \cite{KANG97}. In addition, it was found that after introducing strong pinning by high-energetic Xe$^+$ ion irradiation, a triple sign change evolves \cite{KANG00}. Further investigations concerned the resistivity in magnetic fields oriented perpendicular and parallel to the CuO$_2$ planes, the critical current density, the angular dependence of the depinning field \cite{SALE04b} and the normal and mixed-state Hall effect \cite{SALE04} in partially Re-substituted Hg$_{0.9}$Re$_{0.1}$Ba$_2$CaCu$_2$O$_{6+\delta}$  (HgRe-1212) thin films. The Hall effect's dependence on the angle between the magnetic field and the $ab$ planes in the Hg-1212 thin films could be explained by the common behavior of HTSCs in the normal state and a renormalized superconducting fluctuation model for the temperature region close to $T_c$, where the Hall effect exhibits its first sign change \cite{RICH06}.

Recently, revived interest in the $n = 1$ HBCCO compound has emerged, as it is a model system for a single-CuO$_2$ layer cuprate. Investigations on underdoped samples of this compound \cite{CHAN14,PELC19} have shed light on the nature of the ubiquitous pseudogap in underdoped cuprate superconductors.

In this paper we investigate the structural and the electrical transport properties of three Hg-1212 thin films, fabricated by pulsed-laser deposition on SrTiO$_3$ substrates. {Although we intentionally included a ``bad'' sample with much higher room-temperature resistivity, we find that fundamental superconducting properties, like the critical temperature and the anisotropic Ginzburg-Landau coherence lengths show only little variation in samples with significantly different granularity.}

\section{Sample preparation and characterization}

Thin films of Hg-1212 were fabricated in {three individual runs (each about four months apart) under the same preparation conditions. The labeling of the samples corresponds to the time sequence. The fabrication is a two-step process using polished $5 \times 5\,$mm large (001) SrTiO$_3$ crystal substrates with similar surface roughness.} Firstly, amorphous precursor films were deposited on the using pulsed-laser deposition (PLD) \cite{BAUE00M} with 25~ns KrF excimer laser pulses ($\lambda=248$~nm) with 10~Hz repetition rate. In a second step, the films were annealed in a mercury vapor atmosphere employing the sealed quartz tube technique \cite{YUN95,YUN96,YUN00}. Typically, sintered targets of nominal composition Ba:Ca:Cu = 2:2:3 are employed for laser ablation and precursor films are deposited at room temperature. For Hg-1212 phase formation and for crystallization of films with $c$-axis orientation annealing at high temperature ($800-830^\circ$C) and high vapor pressure (35 bar) is required. HBCCO thin films usually reveal reduced surface quality, phase purity and crystallinity as compared to other HTSC thin films that are grown in a single-step process, like those of YBa$_2$Cu$_3$O$_7$ (YBCO) and Bi$_2$Sr$_2$Ca$_{n-1}$Cu$_n$O$_{2(n+2)+\delta}$ (BSCCO). However, phase-pure epitaxial Hg-1212 films with improved surface morphology are achieved by using mercury-doped targets (Hg:Ba:Ca:Cu $\approx$ 0.8:2:2:3) for laser-ablation and by deposition of precursor films at an elevated substrate temperature $T_S = 350^\circ$C.

\begin{figure}
\begin{center}
\includegraphics[width=\columnwidth]{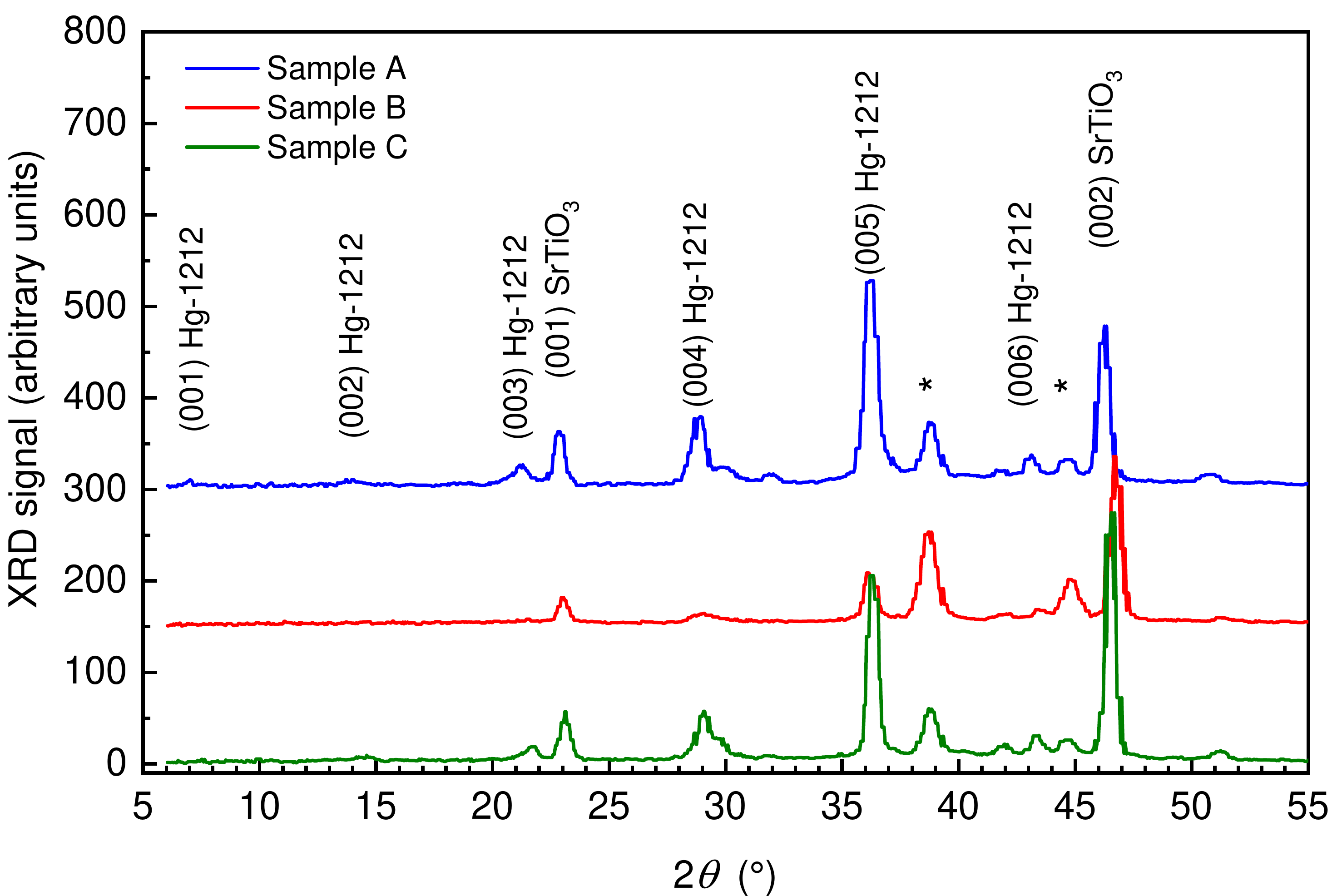}
\end{center}
\caption{X-ray diffraction scan of three HgBa$_2$CaCu$_2$O$_{6+\delta}$ (Hg-1212) thin
films. The curves are offset for better visibility (sample A: top curve, sample C: bottom curve). The (00$l$) indices of Hg-1212 and the SrTiO$_3$ substrate are indicated and  unknown spurious phases are marked by asterisks. Note the significant reduction of the Hg-1212 peak heights of sample B.}
\label{fig_XRD}
\end{figure}

Figure~\ref{fig_XRD} shows the x-ray diffraction (XRD) data of three samples. The (00$l$) indices of Hg-1212 are clearly visible. The full width at half maximum (FWHM) of the (005) rocking curve is about $7^\circ$ for sample A, $6^\circ$ for sample B, and $1^\circ$ for sample C, respectively (data not shown). Besides of the well pronounced peaks resulting from the SrTiO$_3$ substrate only small traces of unknown spurious phases are visible and marked by asterisks. In sample B, however, the {Hg-1212 XRD peaks have lower heights as compared to those of SrTiO$_3$ and the spurious phases}, indicating a smaller amount of the Hg-1212 phase and a larger portion of spurious phases {and voids}.

\begin{figure*}
\includegraphics[width=\textwidth]{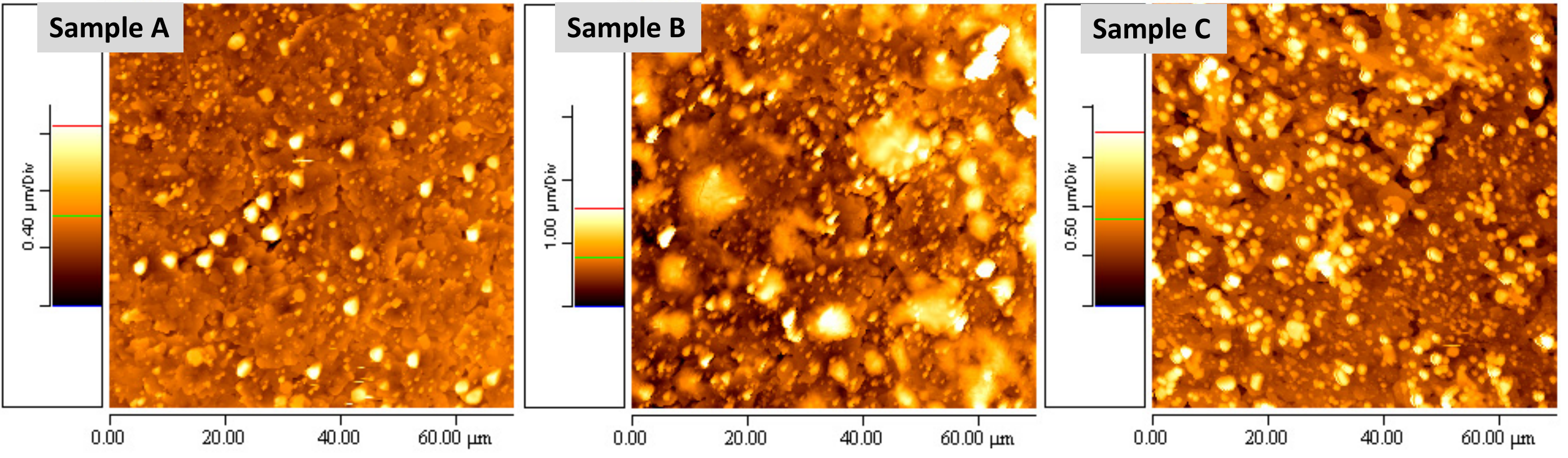}
\caption{Atomic force microscopy height profiles of samples A, B, and C. All pictures display an area of $70 \,\mu\rm{m} \times 70 \,\mu\rm{m}$. Bright spots indicate particulates on the film surface.}
\label{fig_AFM}
\end{figure*}

The surface textures of the three samples are measured by atomic force microscopy (AFM) and are displayed in Fig.~\ref{fig_AFM}. The annealed films reveal a dense and homogeneous structure, an evenly surface without $a$-axis oriented grains, and no regions of unreacted material. Bright spots in the AFM scans indicate particulates of $3 \dots 6\, \mu$m diameter and $0.5 \dots 1\,\mu$m height, which are typically found in HTSC thin films fabricated by PLD and, for instance in YBCO, can be removed by mechanical and chemical polishing \cite{PEDA10}. Sample B exhibits a coarser grain morphology and the larger angular spread of the x-ray rocking curve indicates an enhanced misorientation of the grains.

\section{Experimental setup for the electrical measurements}

For the electrical transport measurements, the Hg-1212 films were patterned by standard photolithography and wet-chemical etching into strips with two pairs of adjacent side arms. Electrical contacts were established by Au wire and silver paste on Au pads that were previously evaporated on the sample's side arms. The thicknesses of the films were determined by AFM. The main parameters of the three samples are summarized in Table~\ref{tab_prop}.

\begin{table*}
\caption{\label{tab_prop}Properties of the Hg-1212 films, where $w$ is the width of the patterned strip, $l$ the distance between the voltage probes, $t_z$ the film thickness, and $T_c$ the critical temperature defined by the inflection point of the resistive superconducting transition curve. {Note that sample A has similar bridge dimensions as the other samples, but different voltage probes have been used.} The intercept of the extrapolated linear normal-state resistivity is $\rho_{xx}(0\,$K), and $\xi_{ab}(0)$ are the in-plane and $\xi_{c}(0)$ the out-of-plane Ginzburg-Landau coherence lengths at $T = 0$, respectively.}
\vskip 10pt
\begin{ruledtabular}
\begin{tabular}{@{}cccccccc}
Sample&$w$&$l$&$t_z$&$T_c$ &$\rho_{xx}(0\,$K)&$\xi_{ab}(0)$&$\xi_{c}(0)$\\
&($\mu$m)& ($\mu$m)&($\mu$m)&(K)&($\mu \Omega$\,cm)&(nm)& (nm)\\
\hline
A&$140 \pm10$&$620 \pm 5$&$0.41 \pm 0.09$&122.0&-36& $2.4 \pm 0.3$&$0.18 \pm 0.05$\\
B&$138 \pm2$&$1240 \pm 5$&$0.46 \pm 0.14$&121.2&-43& $2.3 \pm 0.3$&$0.24 \pm 0.05$\\
C&$128 \pm6$&$1240 \pm 5$&$0.43 \pm 0.19$&121.7&-42&$2.8 \pm 0.2$&$0.17 \pm 0.02$\\
\end{tabular}
\end{ruledtabular}
\end{table*}

Resistivity and Hall effect measurements were performed in a  closed-cycle cryocooler mounted between the pole pieces of an electromagnet. DC currents were provided by a Keithley 2400-LV constant current source and the longitudinal and transverse voltages were recorded simultaneously with the two channels of a Keithley 2182 nanovoltmeter. The directions of both the current and the magnetic field were reversed multiple times for every data point to cancel spurious thermoelectric signals, transverse voltages stemming from contact misalignment, and to enhance the signal to noise ratio. The temperature stability at individual setpoints is better than $\pm$ 0.01\,K.

\section{Results and Discussion}

\begin{figure}
\begin{center}
\includegraphics[width=\columnwidth]{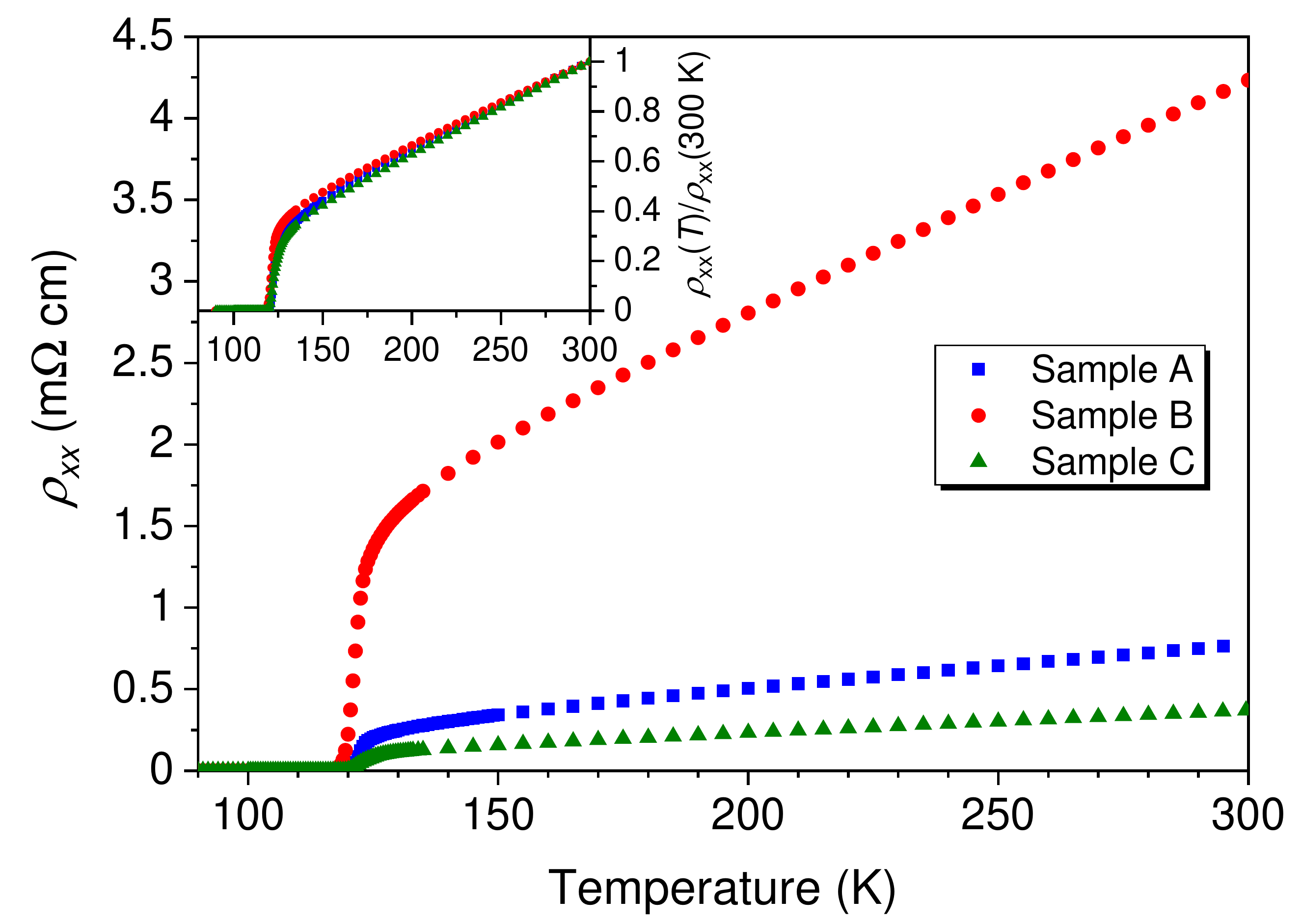}
\end{center}
\caption{Electrical resistivity of the three Hg-1212 films as a function of temperature. The inset shows the resistivities normalized to their values at 300\,K.}
\label{fig_rho}
\end{figure}

The temperature dependencies of the longitudinal resistivity of the three samples are compared in Fig.~\ref{fig_rho}. They show a linear behavior of the normal state resistivity $\rho_{xx}$, typical for optimally-doped cuprate HTSCs, and a reduction of $\rho_{xx}$ below the linear trend above $T_c$ stemming from superconducting fluctuations \cite{LANG94}.

The inset of Fig.~\ref{fig_rho} demonstrates that, despite of the significantly different absolute values of $\rho_{xx}$, all three samples exhibit a similar qualitative behavior when a scaling according to $\rho_{xx}(T)/\rho_{xx}(300\,$K) is applied. This fact contrasts with the typical observation in HTSCs with point defects like disordered oxygen atoms \cite{WANG95b,LANG10R}, where such a scaling is violated and the intercept $\rho_{xx}(0\,$K) from an extrapolation of the normal state is shifted to higher values when the concentration of point defects is increased. In fact, $\rho_{xx}(0\,$K) is slightly negative and similar for all three samples, indicating a minor influence of point defects on the resistivity. Remarkably, the critical temperatures of all samples are similar, too, as listed in  Table~\ref{tab_prop}. These observations indicate that the intragrain resistivities of the samples are similar, but different granularity and different amount of voids and bad-conducting spurious phases lead to a large variation of the macroscopic resistivity.

An analysis of thermodynamic superconducting order parameter fluctuations (SCOPF) \cite{LARK05M} allows one to determine the anisotropic Ginzburg-Landau coherence lengths in superconductors. This method has been applied to many HTSCs but only rarely to Hg-1212 and is based on an evaluation of the paraconductivity $\Delta \sigma_{xx}(0)$ in zero and the paraconductivity $\Delta \sigma_{xx}(B_z)$ in a moderate magnetic field $B_z$, applied perpendicular to the crystallographic $ab$ plane. The total measured conductivity is $\sigma_{xx}(0) = \sigma_{xx}^N(0) + \Delta \sigma_{xx}(0)$ and $\sigma_{xx}(B_z) = \sigma_{xx}^N(B_z) + \Delta \sigma_{xx}(B_z)$, respectively. Commonly, the normal-state conductivities $\sigma_{xx}^N(0)$ and $\sigma_{xx}^N(B_z)$ are determined by extrapolating the linear temperature dependence of the resistivity in the normal state towards lower temperatures. In our samples no deviations from a linear trend at temperatures  between 200~K and 300~K are noticeable and, hence, it is assumed that the SCOPF are negligible above 200~K and furthermore no influence of a pseudogap behavior is expected \cite{SOLO09R}.

In a magnetic field, however, such a procedure requires the assumption that the normal-state magnetoresistance is negligible, i.e., $\sigma_{xx}^N(B_z) \simeq \sigma_{xx}^N(0)$ in the temperature range under investigation \cite{SEKI95}. In fact, the parameters for the linear fits to $\sigma_{xx}(B_z)$ and $\sigma_{xx}(0)$ above 200~K are the same, which indicates a negligible normal-state magnetoresistance in all samples.

Several different processes contribute to SCOPF \cite{LARK05M}, but under the conditions explored in this work, the Aslamazov-Larkin (AL) \cite{ASLA68} process dominates by far. The Lawrence-Doniach (LD) model \cite{LAWR71} is an appropriate extension for two-dimensional layered superconductors with the out-of-plane Ginzburg-Landau coherence length at $T = 0$, $\xi_c(0)$, as the sole fit parameter. The paraconductivity is given by
\begin{equation}\label{eq_LD}
\Delta \sigma_{xx}^{LD} = \frac{e^2}{16 \hbar d}(1+2 \alpha)^{-1/2} \epsilon^{-1},
\end{equation}
where $e$ is the electron charge, $\hbar$ the reduced Planck
constant, $d = 1.2665$\,nm the distance between adjacent CuO$_2$ double layers \cite{RADA93}, and $\epsilon = \ln(T/
T_c)\approx (T-T_c)/T_c$ is a reduced temperature. The dimensionless coupling parameter between the superconducting layers is $\alpha = 2 \xi_c^2(0) d^{-2} \epsilon^{-1}$.

A magnetic field oriented perpendicular to the CuO$_2$ layers leads to a reduction of SCOPF by orbital and Zeeman pair breaking, which is also reflected by a decrease of the mean-field $T_c$. The Zeeman interaction is important for an orientation of the magnetic field parallel to the CuO$_2$ layers only \cite{LANG95a} and can be neglected in the present analysis. The paraconductivity in finite magnetic field considering the orbital interaction with the AL process (ALO) is \cite{HIKA88}
\begin{widetext}
\begin{equation}\label{eq_ALO}
\Delta \sigma_{xx}^{ALO} = \frac{e^2}{8 \hbar h^2} \int\limits_0^{2 \pi/d} \epsilon_k \left[ \psi \left(\frac{1}{2}+\frac{\epsilon_k}{2h} \right)-\psi \left(1+\frac{\epsilon_k}{2h} \right) + \frac{h}{\epsilon_k} \right] \frac{\textrm{d}k}{2\pi},
\end{equation}
\end{widetext}
where $\epsilon_k = \epsilon [1+\alpha(1-\cos kd)]$, $k$ is the momentum parallel to $B_z$, $\psi$ is the digamma function, and $h = \ln[T_c(0)/T_c(B_z)] = 2e\xi_{ab}^2(0)B_z/\hbar$ reflects the reduction of $T_c$ in the magnetic field.

\begin{figure*}
\includegraphics[width=\textwidth]{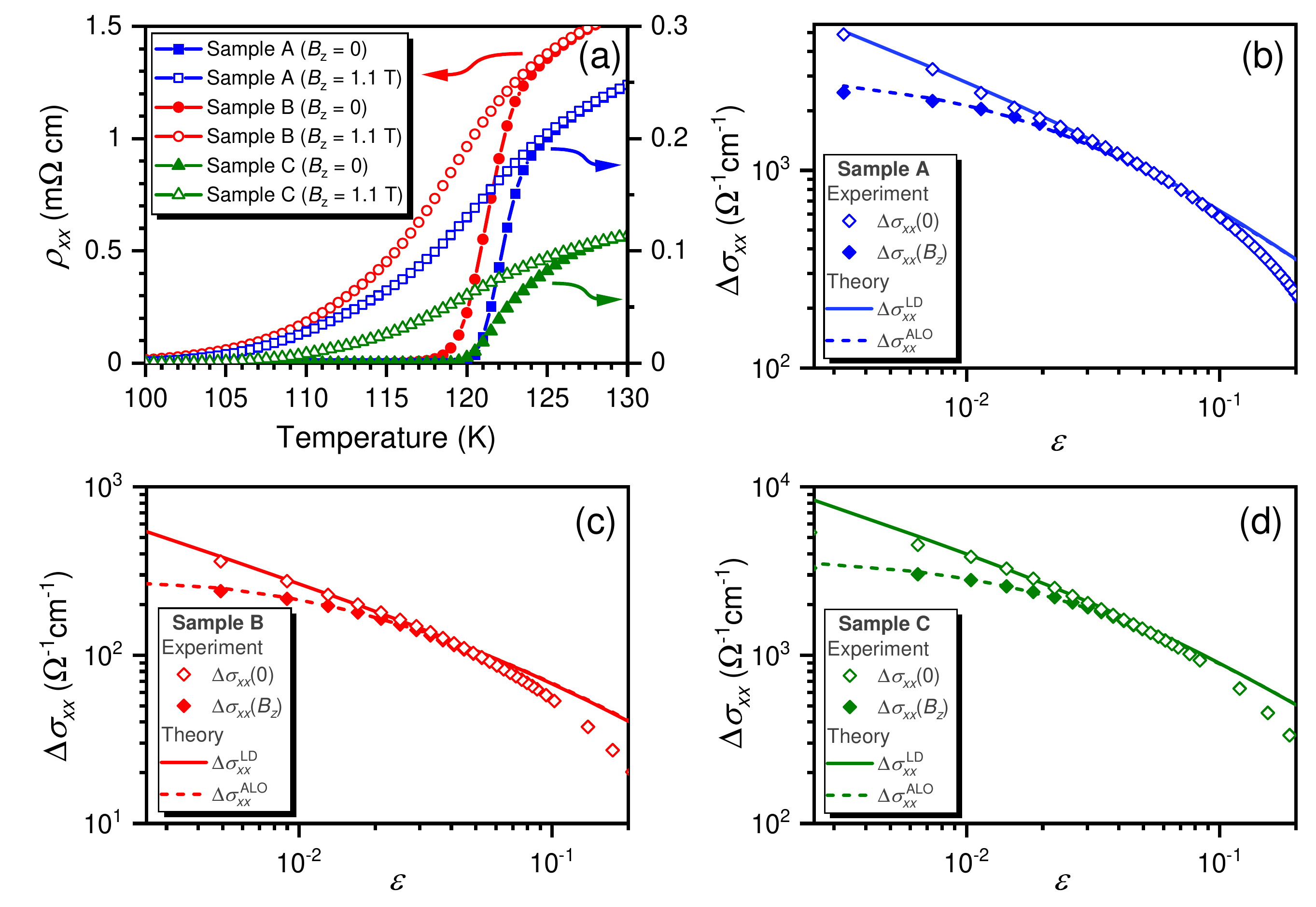}
\caption{(a) Superconducting transition of the three Hg-1212 films in zero and finite magnetic fields. (b -- d)  Paraconductivities of the samples as a function of the reduced temperature $\epsilon = \ln (T/T_c)$ in zero magnetic field and in $B_z = 1.1\,$T. Symbols indicate the experimental data, full lines the fits to Eq.~\ref{eq_LD} for the paraconductivity in zero field and broken lines the fits to  Eq.~\ref{eq_ALO} in finite field, respectively.}
\label{fig_para}
\end{figure*}

Figure~\ref{fig_para}(a) shows the superconducting transitions in zero and finite magnetic fields and Figs.~\ref{fig_para}(b--d) the resulting paraconductivities  $\Delta \sigma_{xx}(0)$ and $\Delta \sigma_{xx}(B_z)$ as a function of the reduced temperature $\epsilon$ together with fits to the LD (Eq.~\ref{eq_LD}) and ALO (Eq.~\ref{eq_ALO}) processes. To account for the higher resistivities observed in samples A and B due to a larger fraction of non-superconducting voids and spurious phases, the theoretical paraconductivity curves are scaled by a factor of 0.70 (0.09) for sample A (sample B) that is determined during the fit procedure. Of course, this introduces an additional uncertainty for the evaluation of $\xi_c(0)$ in samples A and B but has only a minor impact on $\xi_{ab}(0)$.

In paraconductivity studies of HTSCs three temperature regions can be distinguished. Very close to $T_c$ the fluctuating superconducting domains start to overlap and the paraconductivity falls below the predictions of Eqs.~(\ref{eq_LD}) and (\ref{eq_ALO}). This situation is accounted for by renormalized fluctuation theories \cite{IKED89,ULLA91}, which do not contribute to the determination of the coherence lengths and hence are outside of the scope of the present analysis. A counteracting effect can be evoked by a non-homogenous $T_c$ \cite{LANG94c}. Both corrections, as well as the exact value of the mean-field $T_c$ used for the calculation of the reduced temperature $\epsilon$ are relevant for $\epsilon < 0.01$ only. On the other hand, a high energy cutoff of the fluctuation spectrum \cite{CARB01} leads to a smaller paraconductivity as compared to theory for $\epsilon > 0.1$ that can be modeled by a heuristic function resulting in a very similar $\xi_c(0)$ \cite{LERI01}. In the intermediate temperature region $0.01 < \epsilon < 0.1$ the fit parameters can be determined with good precision.

The resulting values of the coherence lengths are listed in Table~\ref{tab_prop}. {Note that despite of the large variation of the resistivities of the samples, their coherence lengths are quite similar. Still, a systematic trend can be presumed.} With degradation of the morphology and increase of the resistivity, $\xi_c(0)$ increases, while $\xi_{ab}(0)$ decreases, leading to a reduction of the anisotropy in the superconducting state. {An increase of crystallographic misorientation between individual grains naturally leads to a reduction of the anisotropy, which is here determined averaged over the entire film, and is also evidenced by the broader rocking curves of samples A and B.}

Compared to other studies in grain-aligned polycrystalline Hg-1212 samples we find about half as long out-of-plane coherence lengths \cite{PUZN95} and larger \cite{PUZN95,THOM96a} or similar \cite{HUAN94} in-plane coherence lengths, indicating a higher anisotropy $\gamma = \xi_{ab}(0)/\xi_{c}(0) \sim 9.6 \dots 16.5$ in our samples. In a HgRe-1212 thin film $\gamma \sim 7.7$ was estimated \cite{SALE04b}, while higher values $\gamma \sim 29$ \cite{BRAS96} and $\gamma \sim 52 $ \cite{VULC96} were reported in HgBa$_2$CuO$_{4+\delta}$ and HgBa$_2$Ca$_2$Ca$_2$Cu$_3$O$_{8+\delta}$ single crystals, respectively. These findings point to a correlation between sample morphology and measured anisotropy.

\begin{figure*}
\centering
\begin{minipage}{.5\textwidth}
\centering
\includegraphics[width=\linewidth]{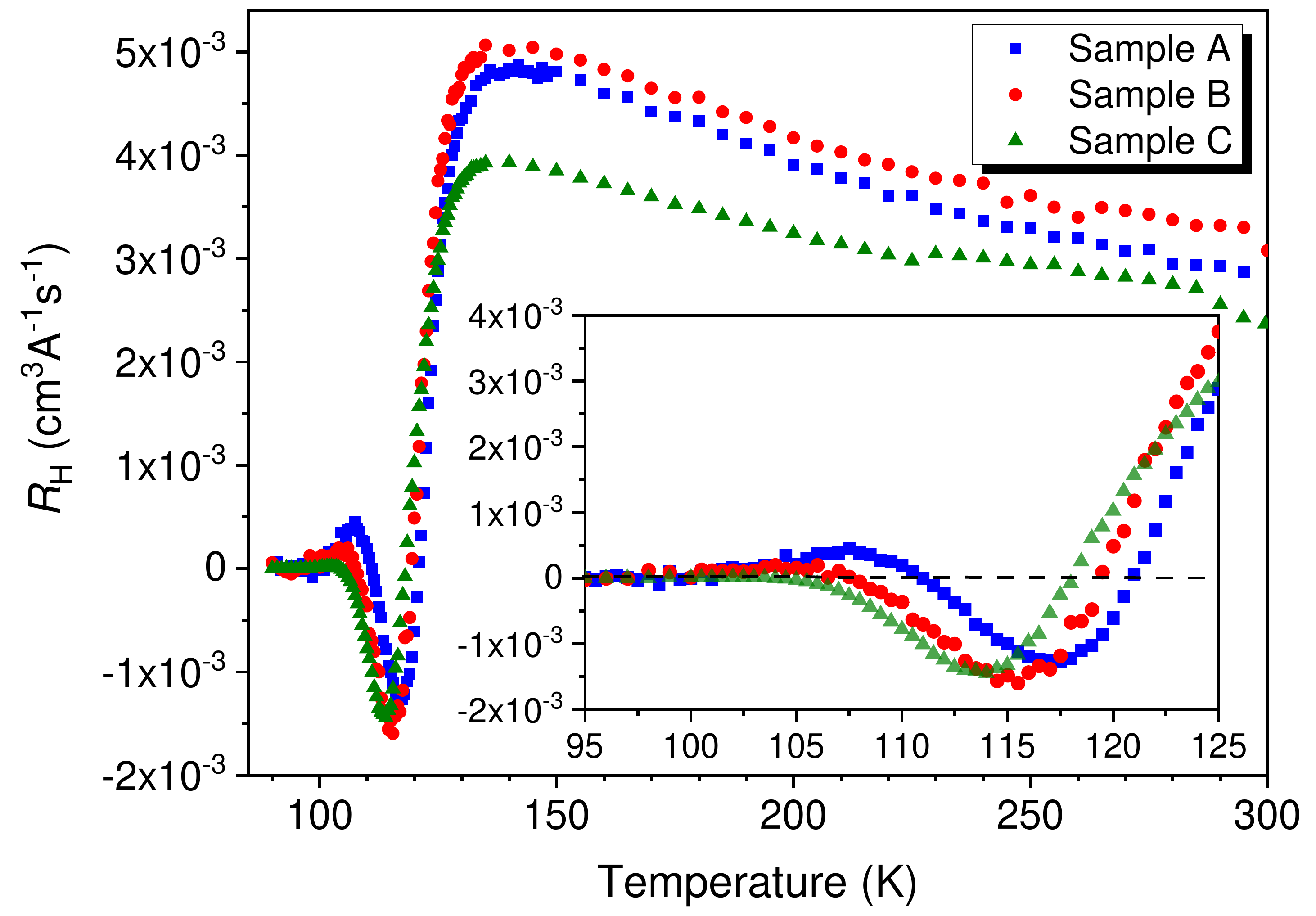}
\end{minipage}%
\begin{minipage}{.5\textwidth}
\centering
\includegraphics[width=\linewidth]{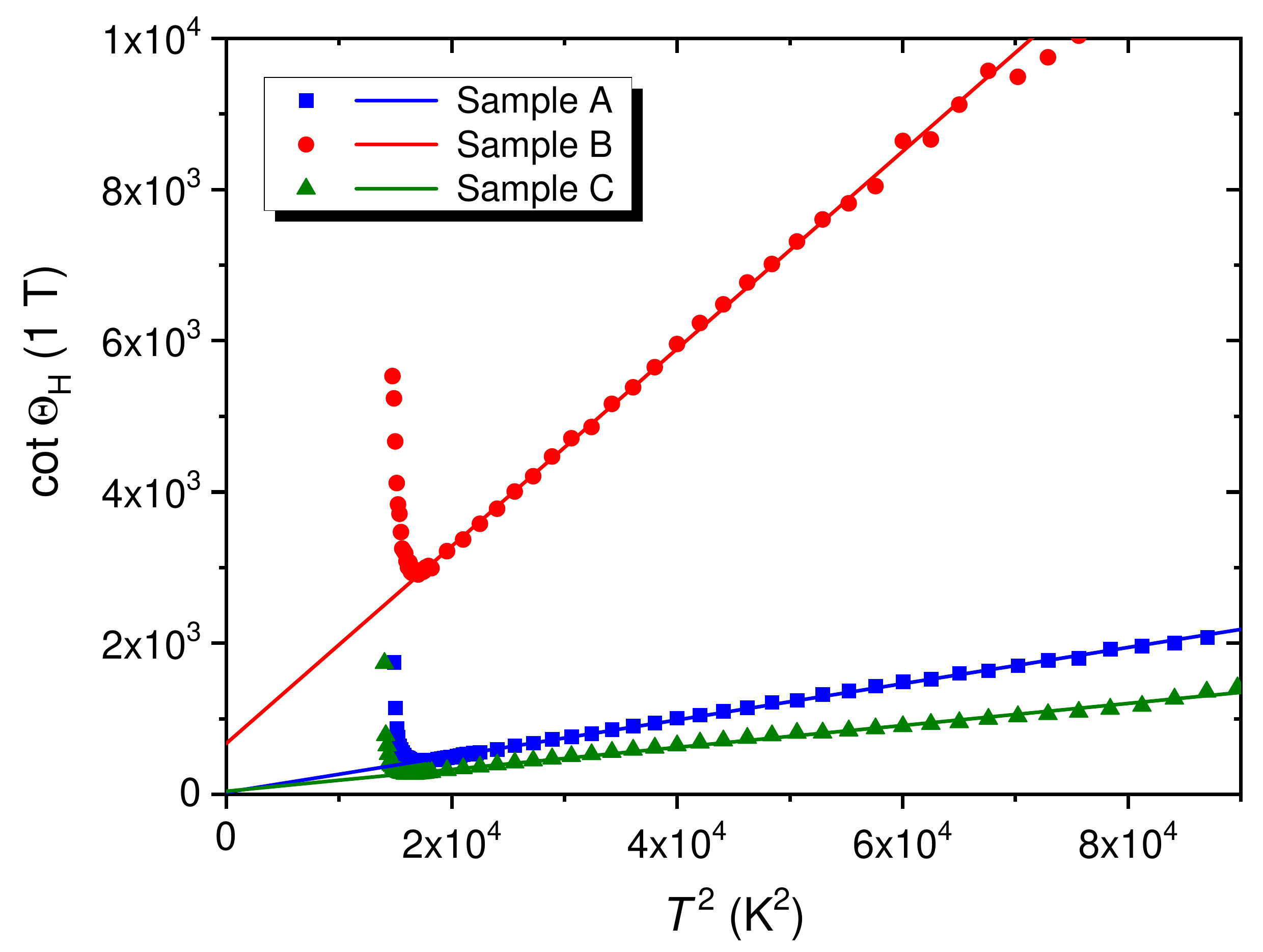}
\end{minipage}
\caption{Hall effect in the three Hg-1212 films, measured at $B_z = 1.1$\,T. Left panel: Hall coefficient $R_H$ as a function of temperature. The inset shows a blow-up of the temperature region around $T_c$. Right panel: Cotangent of the Hall angle $\Theta_H$ normalized to 1~T as a function of the square of the temperature. Only data above $T_c$ are shown.  The lines indicate fits to the data in the normal state.}
\label{fig_Hall}
\end{figure*}

The Hall coefficient is $R_H = E_{y}/(j_{x} B_z)$, where $E_{y}$ is the transverse electric field measured between adjacent side arms of the sample, $j_{x}$ the current density along the strip-shaped Hg-1212 film, and $B_z$ the magnetic field perpendicular to the film surface. Remarkably, $R_H$ is similar in all samples, as can be noticed in Fig.~\ref{fig_Hall}(left panel).

In the normal state $R_H$ is positive (hole-like) and increases towards lower temperatures, followed by a sharp drop around $T_c$ and a subsequent change to negative values. Furthermore, a second sign reversal back to positive $R_H$ is noticeable, but differently pronounced in the samples as displayed in the inset of Fig.~\ref{fig_Hall}(left panel). Qualitatively, our observations are in line with previous investigations in Hg-1212 \cite{KANG97,SALE04} and Bi-2212 films \cite{LANG95,LANG01}. The peculiar sign change of $R_H$ from positive in the normal state to negative in the vortex-liquid regime is still not consensual \cite{AO20,*ZHAO20} and is out of the scope of the present work. Renormalized superconducting fluctuations \cite{PUIC04}, collective vortex effects \cite{AO98}, and pinning centers \cite{KOPN99} are some possible explanations.

While the domain of negative $R_H$ looks similar in all samples, the positive $R_H$ data at the low temperature tail show more variations. {In YBCO, such a double sign reversal is rarely observed and then attributed to vortex-lattice melting {\cite{DANN98}} or pinning at twin boundaries {\cite{GOB00}}. In Hg-1212 it is considered} an intrinsic property \cite{KANG97}, like in Bi-2212, where it becomes more prominent when pinning is reduced in enhanced current densities \cite{LANG01}. But vortex pinning within the grains leads to vanishing of the Hall signal and can lead to a cut-off before the second sign change develops. {One might then speculate that different intragrain pinning properties might cause the differences in the positive low temperature peaks of $R_H$.}

The Hall effect in the normal state is displayed in the right panel of Fig.~\ref{fig_Hall} in an appropriate scaling to demonstrate that, for all three samples, it follows Anderson’s law \cite{ANDE91} $\cot \Theta_H = \alpha T^2 + C$, where $C$ is proportional to the density of carrier scattering defects and $\alpha$ is a measure of the carrier density. The linear trend can be observed in the temperature range from $\sim 135 \,$K to 300\,K, up to higher temperatures than in HgRe-1212 thin films \cite{SALE04b}. The upturn close to $T_c$ is due to the onset of SCOPF. At first sight, $\alpha$ and $C$ appear to be quite different. Since $\cot \Theta_H = \rho_{xx}/(R_H B_z)$,  the voids in the sample that lead to enhanced $\rho_{xx}$, as seen in Fig.~\ref{fig_rho}, in a similar manner effect $\cot \Theta_H$. Hence, if the same scaling as in the inset of Fig.~\ref{fig_rho} is applied to the curves of samples A and B, taking sample C as the reference, a comparison can be made. It turns out that $\alpha$ is similar in all samples, {reflecting a similar carrier density inside the grains, while} the intercept $C$ is largest in sample B and smallest in sample A. {Although only a rough estimate, it} would indicate the smallest density of carrier scattering defects {inside the grains} in sample A and be consistent with the previous observation that samples A exhibits the weakest intragrain vortex pinning.

Finally, we discuss the minor variation with sample morphology of $R_H$ observed in Fig.~\ref{fig_Hall}(left panel) to the contrasting large spread of the resistivities (see Fig.~\ref{fig_rho}). Volger \cite{VOLG50} has theoretically considered a material consisting of well conducting grains separated by thin layers of lower conductivity, which in our samples can be attributed {tentatively} to grain boundaries, voids, and spurious badly conducting phases. In this scenario, the experimentally determined average resistivity can be dominated by the high-resistance domains and can thus be much higher than the intragrain resistivity. On the other hand, the experimentally found $R_H$ will not be very different from its intragrain value. {Note, that the narrow sample dependence of $R_H$ at temperatures slightly above $T_c$ is comparable to the relative variation of the superconducting coherence lengths, which also represent intragrain properties.}

Alternatively, one could consider that the voids in the material reduce the cross section of the current path. Then, using the macroscopic dimensions of the sample for the calculation, the resistivity will be overestimated. However, the local current density in the grains is larger than its average value, giving rise to an enhanced transverse Hall voltage. Since the Hall voltage is probed quasi-electrostatically, \emph{intergranular} resistances are negligible and the Hall voltages of individual grains add up in a series connection across the width of the thin film. Intuitively, this is also reflected by the calculation of $R_H$ into which only the film thickness enters, whereas for the evaluation of the resistivity the sample's thickness, width, and the probe distance are relevant.

\section{Conclusions}
In summary, we have investigated the resistivity and the Hall effect in three Hg-1212 thin films of different morphologies, which were characterized by x-ray diffraction and AFM scans. Despite of a large variation of the absolute values, the resistivity of all samples is linear in the normal state, as it is observed in optimally doped HTSCs. The critical temperatures $T_c \sim 121.2\,$K $\dots 122.0\,$K are similar in all samples{, too,} and the deviations from the linear resistivity trend due to SCOPF allow for the determination of the in-plane $\xi_{ab}(0) \sim 2.3\,$nm\,$\dots 2.8\,$nm and out-of-plane  $\xi_{c}(0) \sim 0.17\,$nm\,$\dots 0.24\,$nm Ginzburg-Landau coherence lengths.

{In sharp contrast to the resistivity,} the normal-state Hall effect is similar in the three samples and is dominated by their intragranular properties. It allows to conclude {that inside the grains the carrier density is almost the same in all samples, but the density of carrier scattering defects is different.} The Hall effect in the superconducting state exhibits two sign changes, from which the one at lower temperatures is sample dependent and can indicate different vortex pinning properties {due to different defect densities inside the grains. Finally, our analyses of various transport measurements on different samples indicate that the intragranular intrinsic properties of the Hg-1212 films can be estimated adequately despite of their diverse macroscopic resistivities.}

\section*{Acknowledgements}
This work was supported by the Austrian Science Fund under grants I4865-N and P18320-N07 and the COST Actions CA16218 (NANOCOHYBRI) and CA19108 (HiSCALE) of the European Cooperation in Science and Technology.

\bibliography{C:/Users/Lang/CloudStation/docs/Lib/Biblio/Vip}

\end{document}